# Scaling Relations for self-Similar Structures and the Cosmological Constant

C.Sivaram
Indian Institute of Astrophysics
Bangalore,India

**Abstract:** Scaling relations for the mass, angular momentum and other properties of a wide range of self-similar structures in the universe are seen to have universal features. As a consequence of the ideas elaborated in earlier papers these relations can be connected to a background constant curvature given by the cosmological constant dominating cosmical dynamics.

In recent papers [13-18], it was pointed out that the surface gravities of a whole hierarchy of astronomical objects (i.e. globular clusters, galaxies, clusters, super clusters, GMC's etc.) are more or less given by a universal value $a_o \cong cH_o \cong 10^{-8}\,\text{cms}^{-2}$. Thus

$$a = \frac{GM}{R^2} \cong a_o, \tag{1}$$

for all these objects, M being their typical mass and R their typical radius. Also interestingly enough it was also pointed out [4-7] that the gravitational self energy of a typical elementary particle (hadron) was shown to be $E_G \cong \frac{Gm^3c}{\hbar} \cong \hbar H_o$ implying the surface gravity for the particle of

$$a_h \cong \frac{GM}{r^2} \cong \frac{Gm^3c}{\hbar}.\frac{c}{\hbar} \cong cH_o \cong a_o \tag{2}$$

the same as eq.(1).

It was further noted in the above papers, that eqs.(1) and (2) imply that the gravitational self energy densities of all these objects should then have the same universal value, i.e.

$$\rho_G \cong \frac{GM^2}{8\pi R^4} \tag{3}$$

is the same for all the above objects and this was shown to be equal to the critical closure density of the universe $\rho_c \cong 10^{-29}\text{gm.cm}^{-3}$. It was again pointed out that for a vacuum dominated universe, i.e. for a cosmological constant ($\Lambda$) dominated universe

$$\rho_c \cong \frac{\Lambda c^4}{8\pi G} \tag{4}$$

Thus relation: [16, 17]

$$(\rho_G)_{particle} = (\rho_G)_{gal} = (\rho_G)_{GC} = (\rho_G)_{\sup ercluster} \cong \rho_c \tag{5}$$

expresses the remarkable result that the gravitational self energy density of an elementary particle, a galaxy, a galactic cluster or a super cluster (also GC's, GMC's) are all the same ($\rho_G$) and equal to the critical cosmological matter density($\rho_c$) which for a cosmological constant dominant dynamics equals the background vacuum energy density. It was hinted in the above papers, that this was to be taken as a kind of cosmological paradigm. One can also consider that all of the above hierarchy of objects are autonomous systems and for this their gravitational(binding)self energy density must at least equal or exceed the background cosmological gravitational self energy density which again equals the critical matter density ($(\rho_G)_{univ} \cong \rho_c$), i.e. to be autonomous stable systems, they are required to be gravitationally bound and for this their self gravitational energy density must at least equal the background ambient value. Eqs (3), (4) and (5) would imply that for all of the above hieararchy of objects.

$$\frac{M}{R^2} \cong \frac{c^2}{G}\sqrt{\Lambda} \quad (6)$$

As Λ is constant throught out the expansion of the universe (this is an advantage of considering Λ rather than $H_o$, and its present dominance of $\rho_c$ would be feature of the present epoch), we have

$$\frac{M}{R^2} = constant \cong \frac{c^2}{G}\sqrt{\Lambda} \quad (7)$$

Eq(4) would imply a $\Lambda \cong 10^{-56} cm^{-2}$(this would be the background curvature), so that $M/R^2$is of the order of unity(as $c^2/G\sqrt{\Lambda} \sim 10^{28}10^{-28} \sim 1$ )
Thus:

$$M \; \alpha \; R^2 \quad (8)$$

for the whole range of the above classes of objects. Thus $R_{Universe} \cong 10^{28}\, cm$,
$M_{univ.} \sim 10^{56}$gms; $R_{Gal} \sim 3\times 10^{22}$cms, $M_{gal} \sim 10^{45}$gm, $M_{particle} \sim 10^{-24}$g, $R_{particle} \sim 10^{-12}$cms, etc.

Again the interstellar medium (ISM) is known to be composed of a hierarchy of structures with masses from 1 $M_{sun}$ to $10^6$ $M_{sun}$ and sizes ranging from 10 A.U to $10^2$pc. Again these structures are also more less seen to obey a $M \alpha R^2$ relation. If we consider the solar system extent of $\sim 10^{16}$ cms, here again, $M/R^2 \sim 1$.

Thus to illustrate the relation:

$$\frac{M}{R^2} \cong \frac{c^2}{G}\sqrt{\Lambda} \sim 1,$$

we have some typical examples:

| $\frac{M}{R^2}(\frac{g}{cm^2})$ | Object |
|---|---|
| $\frac{10^{45}}{(3\times10^{22})^2} \sim 1$ | Galaxy |
| $\frac{10^{56}}{(10^{28})^2} \sim 1$ | Universe |
| $\frac{10^{38}}{(10^{19})^2} \sim 1$ | glob.cluster, GMC,etc. |
| $\frac{few\times10^{48}}{(few\times10^{24})^2} \sim 1$ | supercluster |
| $\frac{10^{-23}}{(10^{-12})^2} \sim 1$ | nuclei |
| $\frac{10^{-27}}{(10^{-13})^2} \sim 1$ | Electron |
| $\frac{10^{33}}{(10^{16})^2} \sim 1$ | Solar system, planetary nebula |

Again we can relate the densities and radii of all the above objects, i.e. autonomous structures by the relation

$$\rho R = const \cong 1 \quad (9)$$

as this follows from eqs. (6-8) above.( $\rho \sim \frac{M}{R^2}.\frac{1}{R}$ = const. $\frac{1}{R}$ , . $\rho$ R=const).
(Thus $\rho_{univ.}R_{univ} \sim 10^{-29}10^{29} \sim 1$, $\rho_{gal}R_{gal} \sim 10^{-23}10^{23} \sim 1$, $\rho_h R_h \sim 10^{13}10^{-13} \sim 1$ etc.)

$$\text{i.e. } \rho R = \frac{M}{R^2} \cong \frac{c^2}{G}\sqrt{\Lambda} \sim 1 \qquad (10)$$

(we make the important point that all the above structures or systems, although they are autonomous units, have very low surface gravity, i.e. they are barely bound with a gravitational energy density equal to the background cosmological energy density as explained above, compact objects with large surface gravities like stars, planets, black holes *do not obey the above relations* and would be briefly discussed later).

The above relations, relate mass and radius of a very wide range of structures. Thus $R_{univ.}/R_{particles} \sim 10^{40}$, $M_{universe}/M_{particle} \sim 10^{80}$, $R_{universe}/R_{gal} \sim 10^{6}$, $M_{universe}/M_{gal} \sim 10^{12}$,etc.

It may be argued that some of these objects for eg. galaxies could have a range of masses. For instance there could be a galaxy with a mass ~ $10^8$ $M_{sun}$, four orders smaller than the assumed one. But then the radius is also correspondingly smaller**,** the eq(8) suggesting that it is almost two orders smaller. This suggests that $M/R^2$ which is the quantity of interest remains more or less the same, i.e eqs (7), (10), etc. still hold. Same thing holds for other classes of objects. A planetary nebula with a three times larger M would have a R, 1.5 times smaller, so $M/R^2$ tends to approach the same value. Same argument holds for star forming clouds. Again radioastronomy line observations have indicated that the ISM is composed of a heirarchy of structures with masses from ~ 1 $M_{sun}$ to $10^6$ $M_{sun}$ and the accumulation of observations at many scales reveal a power-law relationship between size R and mass M [1,2] of the form:

$$M \alpha R^d \quad \text{(with d close to 2)}$$

A self gravitational mean field theoritical approach (Landau Lifshitz, 1996), gives d= 2. So a $10^6$ $M_{sun}$ structure would have a size about 50 pc and $M/R^2$ remains more or less the same as given in above relations (eqs.(7) (10), etc.). What is striking is that the relations hold for such a diverse class or hierarchy of objects, from the universe right up to the nuclei.

There was of course no a priori reason to expect this, over such a range of scales. It is not claimed that it will hold for each and every object in the universe, for eg. tidal interactions and other close gravitational encounters could alter somewhat these relations.(A follow up paper would try to give a deeper basis for this).

Let us now come to angular momentum (*J*) or rotational spin of all of the above objects. As shown in Sivaram [8-11, 17] while M and R cover wide range, the rotational velocity V is more or less the same,i.e. has a much smaller range.
Thus:

$$J \sim MVR \;\; \alpha \;\; MR \;\alpha\; R^3 \qquad (11)$$
$$(\text{As } M \;\alpha\; R^2, \text{ from eqs } (6 - 8))$$

So

$$\frac{J}{R^3} = const \qquad (12)$$

This supports the result in Sivaram [11] that the rotational spin density is constant for the whole of the above range of objects.

As $M^2$ $\alpha$ $R^4$ and $\rho$ ~ 1/R (eq. (9)); we also have the scaling relation

$$J \sim \rho M^2 \tag{13}$$

If $m_p$ is the proton mass, $M_{gal} \sim 10^{69}. m_p$

$$\frac{\rho_{gal}}{\rho_{proton}} \sim 10^{-38} \quad \text{as} \quad \rho_{gal} \sim 10^{-24} \text{ gm cm}^{-3},$$

$$\rho_{nucl} \sim 10^{14} \text{gm cm}^{-3} \text{ so that } J_{gal} \sim 10^{138}. 10^{-38} (\hbar/2) \sim 10^{100}\hbar \tag{14}$$

($\frac{\hbar}{2}$) being the proton spin ). This gives the typical galaxy angular momentum of ~ $10^{74}$gcm$^2$sec$^{-1}$ ~ $J_{gal}$

As eq.(12) implies, the rotational spin density or angular momentum density is the same for the proton as well as for a galaxy and is $J/R^3 \cong$ constant ~ $10^9$g.cm$^{-1}$sec$^{-1}$. Denoting $J/R^3$ as $\sigma$, the spin density, which from eq.(12) is constant for a whole range of entities, we can relate $\sigma$ following the Einstein-Cartan(E-C)theory to the background torsion of space-time as: [17, 18]

$$Q = \frac{4\pi G}{c^3}\sigma = \frac{G}{c^3}.\frac{J}{R^3} \tag{15}$$

($Q^i_{jk}$ the torsion tensor is the asymetric part of the connection, ie. $Q^i_{jk} = [\Gamma^i_{[jk],}]$ the symmetric part being the usual christoffel symbols). Just as in Einstein's theory, the energy density is related to the background curvature as $K \cong G\rho/c^2$ , in the E-C theory, the spin density is related to the torsion Q which is the appropriate geometric quantity. Q has the dimension of inverse length, K the curvature being the inverse length squared. Thus it is remarkable that $M/R^2$ and $J/R^3$ are more or less the same for all of the above structures which cover a very wide range in M, R and J. As Q is in a sense, a square root of curvature K, a natural choice for Q, to make eqs. (4), (6), (7) and (14) comapatible is $Q \cong \sqrt{\Lambda}$ .

$$\sigma \cong \frac{J}{R^3} \cong \sqrt{\Lambda}.\frac{c^3}{G} \tag{16}$$

With $\Lambda \cong 10^{-56}$cm$^{-2}$ as before, this would imply universality of $\sigma = \frac{J}{R^3}$ and $\frac{M}{R^2}$ (eg.(7)). Thus given the length scale of a structure, eqs(7) and (15) would enable its mass and rotational spin to be deduced as $M/R^2$ and $J/R^3$, being the same for all of them.

The above relation would also apply for instance to objects of around a few solar masses but with sizes of several hundred A.U like planetary nebulae, collapsing star forming clouds, etc. However for condensed or compact objects with large surface gravity like stars, etc. we have different relations. In going from stars to galaxies, we have $\rho \propto \frac{1}{R^2}$ , M $\alpha$ R, and J $\alpha$ $M^2$ (for the explanation see ref [11]).

Thus $M_{gal}/M_{sun} \sim R_{gal}/R_{sun} \sim 10^{12}$, $\rho_{gal}/\rho_{sun} \sim R^2_{sun}/R^2_{gal} \sim 10^{-24}$, $J_{gal}/J_{sun} \sim (M_{gal}/M_{sun})^2 \sim 10^{24}$. ($J_{sun} \sim 10^{76}\hbar$ , $J_{gal} \sim 10^{100}$ $\hbar$ ).

Eqs.(4), (5), (7)and (15), would imply, $\Lambda \; \alpha \quad \rho_c\sqrt{\Lambda} \quad \alpha \;\; \sigma$ or $\sigma^2 \; \alpha \; \rho_c$, i.e., $(G\sigma/c^3) \cong (G\rho_c/c^2)^{1/2}$, $\cong \sqrt{\Lambda}$ being constant. Also from eq.(10), $\sqrt{\Lambda} \cong \frac{G\rho R}{c^2}$, for the whole range of above objects.

$M/R^2$ being constant, implies a constant mass to area ratio, and is suggestive of a bubble energy with tension T, as energy $E= 4\pi R^2T$. So $E/R^2$ being constant, suggests a universality of tension being the same, this surface tension being given by:

$$M/R^2 = \sqrt{\Lambda}\, c^2/G \cong \text{constant}$$

This relation holds as long as the object size(whether GC, galaxy, or supercluster) $R << \frac{1}{\sqrt{\Lambda}}$ ,$\Lambda$ being the background curvature. For the universe as a whole R becomes comparable, to $\frac{1}{\sqrt{\Lambda}}$ ,so thet $M/R = c^2/G$(as $\sqrt{\Lambda} \sim \frac{1}{R}$ ).

The relation $M/R = c^2/G$ also holds for black holes. A black hole being a disconnected world in its own de-sitter universe (in the interior), perhaps, $R_{Sch} \sim \frac{1}{\sqrt{\Lambda_{gal}}}$ (local curvature, corresponding to its vacuum polarisation energy). So if the object is trapped in its own gravitational field the appropriate relation is $M/R \sim c^2/G$.

This relation also holds near planck scales, when gravity is very strong, i.e. $R_{pl} \sim \frac{1}{\sqrt{\Lambda_{pl}}}$ , $M_{pl}/R_{pl} \sim c^2/G$, so for objects in very strong self gravitational fields, $M/R \sim c^2/G$,i.e. their radius corresponds to the curvature of the strong self gravitational field. For application to elementary particles. (see ref [3], [13-15], [17]) For objects with low surface gravity, with length scales significantly smaller than background radius of curvature eqs.(6), (7), (15), were successfully used for a very wide range of objects as we have already seen. A more detailed physical picture of the significance of the above results, as following from statistical mean field theory (and a renormalization group basis as in ref[16, 17] would be explored in a separate publications. Indeed, the energy(mass) per unit area, i.e. surface tension given by eqs. (6)-(7), i.e. $M/R^2 \sim \frac{c^2\sqrt{\Lambda}}{G}$ , has the same numerical value as that used in nuclear physics as the surface tension in the nuclear liquid drop model. Here the same surface tension occurs for a whole hierarchy of a very wide range of celestial objects ranging from globular clusters, GMC's galaxies, and superclusters of galaxies. The consequences of this most intriguing fact would be explored in a separate publication.

Briefely for a nucleus of mass number A and radius r, one binding surface energy can be written as $4\pi r^2(A^{2/3}-1)T$, where T is the surface tension of the nuclear force, i.e. energy per unit area (the nucleus behave like a liquid drop). For the helium nucleus, A=4, so that $A^{2/3} \approx 16^{1/3} \approx 2.5$. So the nuclear binding energy now becomes: (for the helium nucleus) $\Delta E_n = 6\pi R^2 T$ . Now for T, which is essentially the energy per unit area, we have the same value as above:

$$T = \frac{\sqrt{\Lambda}}{G}c^4 \approx 10^{20}\text{ergs/cm}^2 \tag{17}$$

This gives (when substituted in to $\Delta E_n$) for the binding energy of the helium nucleus as $\Delta E_n = 4.5 \times 10^{-5}$ergs.

i.e. $$4\pi\left(\frac{\hbar}{m\pi c}\right)^2 (A^{2/3}-1)\frac{c^4}{G}\sqrt{\Lambda} = \Delta E_n$$

This is precisely the binding energy released in the conversion of hydrogen to helium. So eq.(16) not only gives the surface energy (energy per unit area) of the large scale structure (galaxies,clusters etc.)but also the nuclear surface tension of the atomic nucleus.

Thus the cosmological vacuum energy also seems to fix T for the atomic nucleus providing a connecting link. Many similar relations exist [20-23].